\documentclass[aps,pra,preprint,floatfix,superscriptaddress]{revtex4}  
\usepackage{float}
\usepackage{graphicx}
\usepackage{amssymb}
\usepackage{dsfont}
\usepackage{amsmath}  
\usepackage{epstopdf}
\begin{document}

\def\ket#1{|#1\rangle} 
\def\bra#1{\langle#1|}
\def\av#1{\langle#1\rangle}
\def\dkp#1{\kappa+i(\Delta+#1)}
\def\dkm#1{\kappa-i(\Delta+#1)}
\def\pp{{\prime\prime}}
\def\ppp{{\prime\prime\prime}}
\def\w{\omega}
\def\k{\kappa}
\def\D{\Delta}
\def\wp{\omega^\prime}
\def\wpp{\omega^{\prime\prime}}

\title{Conditional phase gate between two photons through control of the interaction time with a single atom in a cavity.}
\author{Arkan Hassan and Julio Gea-Banacloche}
\affiliation{Department of Physics, University of Arkansas, Fayetteville, AR 72701}

\date{\today}

\begin{abstract}
We show that the simultaneous interaction of two single-photon fields with a single atom in the V configuration can in principle produce a conditional phase gate of arbitrarily high fidelity, for an appropriate choice of the interaction time, as long as the fields con be described by a single temporal mode (as in an optical cavity); this requires a ``gated'' interaction, where, e.g., dynamical coupling techniques could be used to get the fields in and out of the cavity, and a large detuning induced by a strong external field could be used to turn the atom-field interaction on and off at the right times.  With these assumptions, our analysis shows that the largest gate fidelities are obtained for a cavity containing a single atom, and that adding more atoms in effect ``dilutes'' the system's nonlinearity.  We also study how spontaneous emission losses into non-cavity modes degrade the fidelity, and consider as well a couple of alternate atomic level schemes, namely two- and five-level systems.  
\end{abstract}
\maketitle

\section{Introduction and summary}

Single photons are, in many ways, ideal systems to be used as qubits for quantum information processing tasks, but the lack of a direct interaction between photons makes conditional logic operations challenging, particularly in the optical domain, where the single photon-single atom coupling is too weak, in free space, to effectively mediate the interaction.  Nevertheless, the coupling can be substantially increased by placing the atoms in an optical cavity, and a number of schemes involving single atoms (or even ensembles of atoms) in cavities have been proposed \cite{duan,koshino,chuang}, and several also demonstrated experimentally  \cite{reiserer,rempe,stolz}.

A common feature of all these systems, however, is that the interaction is \emph{sequential}:  the two photons involved interact with the cavity-atom system one at a time.  One reason for this is that (as was shown in \cite{kerr}), in general, if two traveling photons interact simultaneously with an ideal $\chi^{(3)}$ medium, in order to get a conditional phase shift, the interaction typically leaves them in a frequency-entangled state whose small overlap with the initial state results in a very low gate fidelity \cite{note1}.  Subsequently it was shown that this would also be the case for a $\chi^{(2)}$ medium, and also if the nonlinear medium was placed in a cavity \cite{bala1,arkan}.  (See the introduction to \cite{brod1} for a detailed history of these and other difficulties in devising passive, deterministic gates for optical photons.)

Eventually, a theoretical solution to the spectral entanglement problem for traveling photons was found to be, essentially, to make the photon-photon interaction nonlocal by spreading it coherently over many sites \cite{brod1,brod2,konyk,sorensen} or over a suitably nonlocal medium \cite{knight,bala}; however, these proposals have not yet been demonstrated experimentally.  An alternative approach that would make use of a different atom-photon coupling has also been suggested very recently \cite{prx}.

Another way to avoid the spectral entanglement problem, while taking advantage of the enhanced coupling provided by an optical cavity, was proposed in \cite{heuck1,heuck2}.  Basically, this would make use of ``dynamical coupling'' techniques (such as discussed in \cite{raymer}) to get the two photons wholly inside the cavity first, then turn on the interaction with the nonlinear medium, turn it off after a suitable time, and then reverse the loading process to get the photons out with negligible wavepacket distortion.  Since while the interaction is taking place, the photons are described by a single temporal mode (i.e., the cavity mode), no spectral entanglement is possible.  

For the interaction, the authors of \cite{heuck1,heuck2} focused primarily on conventional nonlinear materials, which would require extremely high nonlinearities to achieve the desired phase shifts at the single-photon level.  We explore here, instead, the possibility of using a single, V-type atom as the ``nonlinear medium,'' where the photon-atom interaction could, in principle, be turned off at will by, e.g., applying an external field to the atom to induce a large detuning.  We find that, assuming perfect control and ideal (i.e., lossless) conditions, it is possible to find parameters for which a conditional phase (CPHASE) gate with arbitrarily large fidelity can be achieved.  This turns out not to be trivial: as explained in Section II, our result is actually based on the fact that \emph{efficient} rational approximations exist to the irrational frequencies appearing in the state evolution coefficients.  


We need to note, at this point, that the possibility of using a precise control of the interaction time between two photons, each described by a single temporal mode, and an ensemble of $N$ atoms, to effect a CPHASE gate was originally suggested by Ottaviani et al. \cite{ottaviani,rebic}, and in fact it was this work that provided the original motivation for our research.  As we shall show here, however, the scheme only works optimally for a single atom, as adding atoms in effect ``dilutes'' the nonlinearity of the system until, in the limit of very large $N$, it essentially vanishes (i.e., the phase shift for two photons is just twice the phase shift for a single photon).  This, the second main result of our paper, is established in detail for the V-system in Section III.  In section IV, and for completeness, we also explore the suggestion of \cite{ottaviani,rebic} of extending the system to five levels with a couple of strong driving fields, to possibly take advantage of electromagnetically-induced transparency (EIT) to mitigate the loss of fidelity due to spontaneous emission.

As noted above, most of our paper focuses on the V-system, for several reasons: it is a reasonable simplification of the scheme of Ottaviani et al. \cite{ottaviani,rebic} (superior, in fact, if losses are negligible, as shown in Section IV); it has been at the heart of several earlier proposals (e.g., \cite{chuang,brod1,brod2,sorensen}); and it seems especially suited to work with a polarization encoding.  We note, however, in Section V, that our method would work as well with an ordinary two-level atom and a two-rail encoding, using a scheme similar to that proposed in \cite{nysteen}.  This might, therefore, serve as an alternative to the scheme proposed in Section VI.C of \cite{heuck2}.  This and other related issues are further discussed in the Conclusions (Section VI).


\section{C-PHASE gate using a single V-type atom}

\subsection{The lossless case.}
In this section we consider the interaction of one atom in the V configuration (as shown in Figure 1), with two field modes, $a$ and $b$, with different polarizations, which can initially contain at most one photon each.  The single-temporal mode assumption implies that the interaction takes place in an optical cavity.  This can, in principle, be set up by first loading the photonic traveling fields into the cavity, using techniques such as described in Refs.~\cite{raymer,heuck1,heuck2}; the frequency-conversion process used to this end needs to preserve the individual polarizations.  The finesse of the cavity at the frequency of the $a$ and $b$ modes is assumed to be essentially infinite.  We also assume this frequency to be far detuned from the initial atomic transition frequency, so no interaction between the atom and the fields takes place during the loading process. At $t=0$ we assume the atom is brought into (near) resonance with the fields, and hence allowed to interact with them for $0<t<T$.  At $T$ the interaction is stopped, by again bringing the atom far from resonance, and the fields are extracted from the cavity by the reverse of the frequency-conversion process used to load them in.
\begin{figure}[h]
    \includegraphics[width=8cm]{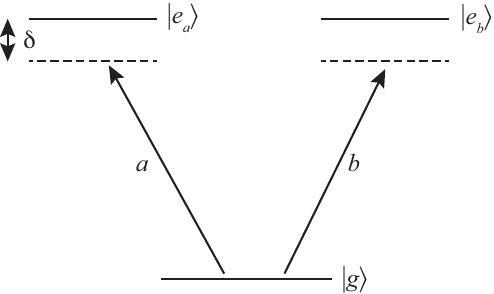}%
    {\caption{The symmetric V configuration.  Fields $a$ and $b$ (same frequency, different polarization) connect the ground state $\ket g$ to the excited states $\ket{e_a}$ and $\ket{e_b}$, respectively.}}
\label{fig:fig1}
\end{figure} 

We assume a symmetric arrangement with equal detunings and coupling constants for both modes, for simplicity; the problem is analytically solvable also without these assumptions, and we have verified that the best results (i.e., highest gate fidelities) are obtained under these conditions.  The Hamiltonian, in an interaction picture for the fields, is then
\begin{equation}
H = \hbar\delta\ket{e_a}\bra{e_a} + \hbar\delta\ket{e_b}\bra{e_b} + \hbar g \bigl(\ket{e_a}\bra g a + \ket{e_b}\bra g b + H.c.\bigr)
\label{e1}
\end{equation}
We have made the dipole and rotating-wave approximations, which are standard for cavity QED in the optical domain.

With one photon initially in each mode and the atom starting in the ground state $|g\rangle$, the system's state at any later time can be written as
\begin{equation}
\ket{\Psi\left(t\right)} = C_{ea}^{(2)} |01\rangle \ket{e_a} + C_{eb}^{(2)} |10\rangle \ket{e_b} + C_{g}^{(2)} |11\rangle |g\rangle
\label{e2}
\end{equation}
leading to the equations of motion
\begin{align}
\dot{C}_{ea}^{(2)} &=  - i \delta {C}_{ea}^{(2)} - i g {C}_{g}^{(2)} \cr
\dot{C}_{eb}^{(2)} &= -i \delta {C}_{eb}^{(2)} - i g {C}_{g}^{(2)} \cr
\dot{C}_{g}^{(2)}  &= -i g  {C}_{ea}^{(2)} - i g {C}_{eb}^{(2)} 
\label{e3} 
\end{align}
In the case only one of the photons, say, the $a$ photon, is initially present, the relevant equations would be instead
\begin{align}
\dot{C}_{ea}^{(1)} &=  - i \delta {C}_{ea}^{(1)} - i g {C}_{g}^{(1)} \cr
\dot{C}_{g}^{(1)}  &= -i g  {C}_{ea}^{(1)} 
\label{e4} 
\end{align}
These equations are easily solved, with the results, for the single-photon case, 
\begin{align}
C_{g}^{(1)}\left(t\right) &=   e^{-i\delta t/2} \left[\frac{i\delta}{2\omega_1} \sin{\omega_1 t} + \cos{\omega_1 t}\right] \cr
C_{ea}^{(1)}\left(t\right) &= -i g e^{-i\delta t/2} \frac{\sin{\omega_1 t}}{\omega_1}
\label{e5}
\end{align} 
and, for the two-photon case,
\begin{align}
C_{g}^{(2)}\left(t\right) &=   e^{-i\delta t/2} \left[\frac{i\delta}{2\omega_2} \sin{\omega_2 t} + \cos{\omega_2 t}\right] \cr
C_{ea}^{(2)}\left(t\right) &= C_{eb}^{(2)}\left(t\right) = -i g e^{-i\delta t/2} \frac{\sin{\omega_2 t}}{\omega_2}
 \label{e6}
\end{align} 
\noindent where 
\begin{align}
&\omega_1 = \frac{1}{2}\sqrt{\delta^{2} + 4 g^{2}}\cr
&\omega_2 = \frac{1}{2}\sqrt{\delta^{2} + 8 g^{2}}
\label {e7}
\end{align}
In order to carry out a successful CPHASE gate on this pair of photons, we would ideally like to choose an interaction time $T$ such that the atom returns to the ground state with unit probability at that time, independently of whether one or both photons are initially present, so that
\begin{equation}
C_g^{(1)}(T) = e^{i\phi_1}, \quad C_g^{(2)}(T) = e^{i\phi_2}
\label{e8}
\end{equation}
and with the phases $\phi_1$ and $\phi_2$ such that 
\begin{equation}
\phi_2 - 2\phi_1 = \pi
\label{e9}
\end{equation}
In what follows we will often refer to $\phi_2 - 2\phi_1$ as the \emph{nonlinear} phase shift, since $\phi_2 = 2\phi_1$ is just what is expected from linear evolution.

While it is not actually possible to satisfy Eqs.~(\ref{e8}) and (\ref{e9}) exactly for a finite interaction time $T$, it is, in principle, possible to get arbitrarily close for a sufficiently long $T$.  This is most easily seen in the resonant case, $\delta =0$, where $C_g^{(1)} = \cos(gt)$ and $C_g^{(2)} = \cos(\sqrt 2 gt)$.  Eqs.~(\ref{e8}) and (\ref{e9}) would then be satisfied if we could have simultaneously $g T=n\pi$, with $n$ an integer (so $\phi_1 =\pi$ or $2\pi$), and $\sqrt 2 g T = m\pi$, with $m$ an odd integer.  This would require $\sqrt 2$ to be of the form $m/n$, i.e., to be a rational number, which it is certainly not; however, a result in number theory \cite{dirichlet} shows that there exist rational approximations to $\sqrt 2$ (and, indeed, to any irrational number) with the property that 
\begin{equation}
\left| \sqrt 2 - \frac m n \right| < \frac{1}{n^2}
\label{e9a}
\end{equation}
with $n$ arbitrarily large.  One then only has to choose one such approximation with odd $m$, and let $gT = n\pi$, to have $|C_g^{(1)}(T)|=1$ and $C_g^{(2)}(T) = \cos(m\pi+\epsilon)$, where $\epsilon \sim 1/n$. Note that the $1/n^2$ bound in (\ref{e9a}) plays an essential role here: if the difference between  $\sqrt 2$ and the fractional approximation  $m/n$ only decreased as $1/n$, then making $gT = n\pi$ would make $\sqrt 2 g T= \sqrt 2 n \pi \sim (m/n+ O(1/n))n\pi = m\pi + \pi O(1)$, and there would be no way to bound the error in $C_g^{(2)}(T)$. 

The best sequence of rational approximations to $\sqrt 2$, as obtained from its partial fraction expansion \cite{ams}, is $1,\, 3/2,\,  7/5, \, 17/12, \, 41/29,\ldots, p_k/q_k,\ldots$, with $p_{k+1} = p_k+2q_k$ and $q_{k+1} = p_k+q_k$ (note that the first of this relations implies that, conveniently, all the numerators are odd).  It is easy to check that the relatively rough approximation  $\sqrt 2 \simeq 17/12$ already yields $C_g^{(2)}(T) = \cos(12 \sqrt 2 \,\pi)=-0.9957$, for $gT=12\pi$.  

Instead of making $C_g^{(1)}(T)$ exactly equal to 1, one might try a slightly different value of $T$ that makes $C_g^{(1)}(T)$ somewhat smaller than 1 but $C_g^{(2)}(T)$ closer to $-1$. A way to quantify, in a single number, the potential impact of these tradeoffs is provided by a \emph{gate fidelity} that can be defined, for this system, in the following way.  Assume the initial state of the atom-field system to be $\ket{\Psi(0)} = (\alpha_{00}\ket{00} +  \alpha_{01}\ket{01} + \alpha_{10}\ket{10} +  \alpha_{11}\ket{11})\ket g$.  Then, ideally, we'd want the state of the field at the time $T$ to be
\begin{equation}
\ket{\Phi_\text{ideal}} =  \alpha_{00}\ket{00} +  \alpha_{01}e^{i\phi_1}\ket{01} + \alpha_{10}e^{i\phi_1}\ket{10} -  \alpha_{11}e^{2i\phi_1}\ket{11}
\label{e10}
\end{equation}
where the phase $\phi_1$ is arbitrary, and given by $e^{i\phi_1} = C^{(1)}_g/|C^{(1)}_g|$; what matters is that Eq.~(\ref{e9}) be satisfied, as indicated by the minus sign in Eq.~(\ref{e10}).  

We can then define the gate fidelity by
\begin{equation}
{\cal F} = \overline{\bra{\Phi_\text{ideal}}\rho_{f}\ket{\Phi_\text{ideal}}}
\label{e11}
\end{equation}
Here $\rho_f = \text{Tr}_{at}\bigl(\ket{\Psi(T)}\bra{\Psi(T)}\bigr)$ is the reduced density operator for the field after tracing over the atomic states, and the overbar means that the quantity $\bra{\Phi_\text{ideal}}\rho_{f}\ket{\Phi_\text{ideal}}$ is to be averaged over the coefficients $\alpha_{ij}$ of the initial state, assuming them to be uniformly distributed in magnitude between zero and 1, with random phases, and satisfying $\sum |\alpha_{ij}|^2 =1$. This formal average yields
\begin{widetext}
\begin{equation}
{\cal F} = \frac{1}{10}\left(1+3|C_g^{(1)}|^2 + |C_e^{(1)}|^2 + |C_g^{(2)}|^2 +|C_e^{(2)}|^2+2|C_g^{(1)}|-\frac{1}{|C_g^{(1)}|^2 }\left(1+2|C_g^{(1)}|\right) \text{Re}\left[\left({ C_g^{(1)}}^\ast\right)^2  C_g^{(2)}\right] \right)
\label{e12}
\end{equation}
\end{widetext}
where $|C_e^{(1)}|^2$  and $|C_e^{(2)}|^2$ are the occupation probabilities of either state $\ket{e_a}$ or  $\ket{e_b}$ in the single- and two-photon cases, respectively.  (Note that, in the lossless case considered in this subsection, one could use $|C_e^{(1)}|^2 + |C_g^{(1)}|^2 =1$ and $2|C_e^{(2)}|^2 + |C_g^{(2)}|^2 =1$ to further simplify the result (\ref{e12}).)


\begin{figure}[h]
    \includegraphics[width=8cm]{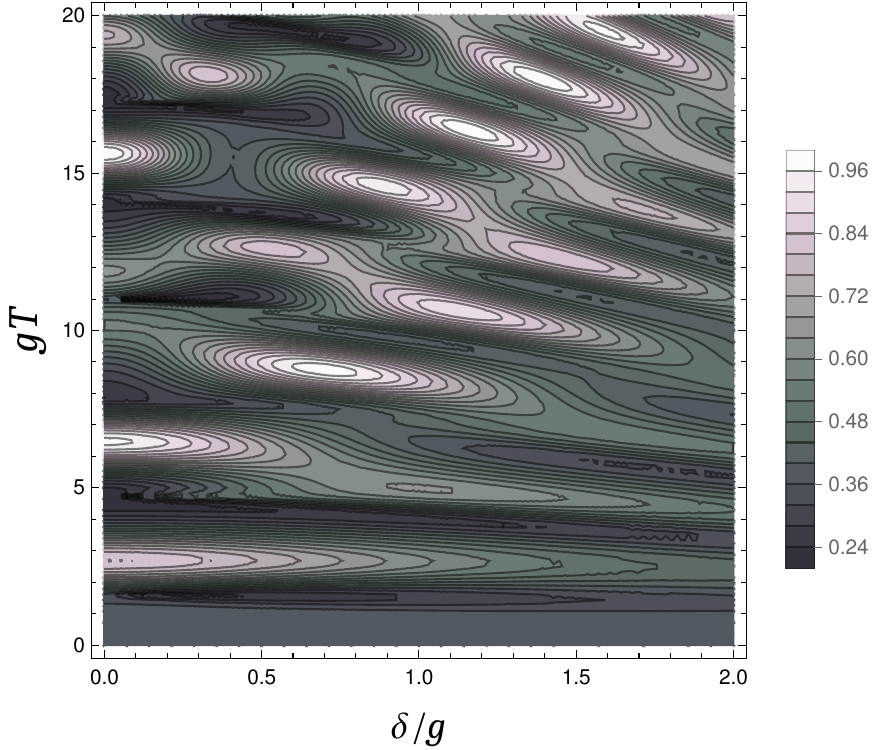}%
    {\caption{Contour plots of the gate fidelity (\ref{e11}) as a function of detuning and interaction time. }}
\end{figure}

Figure 2 shows a contour plot of $\cal F$ as a function of $\delta/g$ and $gT$.  The maxima visible along the vertical axis ($\delta =0$), specifically for $gT=6.473$ and $15.629$, correspond to the first two continued fraction approximations to $\sqrt 2$, namely, $3/2$ and $7/5$, and despite their crudeness they already yield ${\cal F} = 0.9714$ and $0.9950$, respectively.  The next term, $17/12$, mentioned above but not visible in the figure, as it corresponds to the rather large $gT \simeq 12\pi \simeq 37.7$, would yield ${\cal F} = 0.9992$.  (Note that, in a practical application, one could probably not make $gT$ very large, since the effective interaction time $T$ will be limited by losses, as seen in the following subsection.)

Also of potential interest are the fidelity peaks visible in Figure 2 for nonzero detuning $\delta$.  The mathematical explanation for these peaks follows along similar lines to the zero-detuning case.  If, for some $\delta$ and $T$, one can make $\omega_1 T \simeq n\pi$ and $\omega_2 T \simeq m\pi$, then by Eqs.~(\ref{e5}) and (\ref{e6}) one has $C^{(1)}(T) \simeq (-1)^n e^{-i\delta T/2}$ and $C^{(2)}(T) \simeq (-1)^m e^{-i\delta T/2}$.  In that case, $2\phi_1 \simeq -\delta T$ and $\phi_2 \simeq -\delta T/2$ if $m$ is even, or $\phi_2 \simeq -\delta T/2+\pi$ if $m$ is odd.  The conditions (\ref{e8}) and (\ref{e9}) will then be satisfied if one can find three integers, $m,n$ and $q$, such that 
\begin{align}
\frac 1 2 \sqrt{\delta^2 + 4g^2}T &\simeq n\pi \cr
\frac 1 2 \sqrt{\delta^2 + 8g^2}T &\simeq m\pi \cr
\frac 1 2 \delta T &\simeq q\pi
\label{n14}
\end{align}
with $m$ and $q$ of opposite parity.  This requires $n,m$ and $q$ to approximately satisfy 
\begin{equation}
2 n^2 = m^2 + q^2
\label{n15}
\end{equation}
with $m>n>q$.  

All the maxima seen in Figure 2 correspond to approximate solutions of this equation, and it is easy to see that the approximations can be improved, and the gate fidelity along with them, indefinitely, for long enough times, as in the $\delta =0$ case.  For example, letting $n=3q$ Eq.~(\ref{n15}) becomes $17q^2 = m^2$, and one can then choose $m$ and $q$ from the successive optimal approximations to $\sqrt{17}$: $4/1, 33/8, 268/65,\ldots$ \cite{note2}. The first one, $m=4$, $q=1$, when substituted in (\ref{n14}), yields $\delta/g=0.707$ and $gT=8.886$; the actual maximum of $\cal F$ is at $\delta/g=0.699$ and $gT =8.762$, and equals $0.9849$.  (For reference, the largest maximum of $\cal F$ in the region shown in Fig.~2 occurs at $\delta/g = 1.3881$ and $gT = 18.007$, and equals $0.9968$; the corresponding values of $n,m$ and $q$ in Eq.~(\ref{n14}) are $(7,9,4)$.)

Before we proceed to consider the impact of losses, it may be worthwhile to recall here the results we obtained in \cite{arkan} for the same system (two photons incident on a cavity containing a V-system atom) but without the ``gating'' proposed here.  In that case, where the photons are described by wavepackets that simply enter and leave the cavity by transmission through the mirror, and interact with the atom continuously during this process, the maximum gate fidelity achievable (optimizing over all the parameters $g$, $\delta$, $\kappa$, the cavity bandwidth, and $\sigma$, the pulse's spectral width) is ${\cal F} = 0.556$ for a pulse with a Lorentzian spectrum, $f_0(\omega) \propto 1/((\omega-\omega_0)^2+\sigma^2)$.




\subsection{Impact of losses (spontaneous emission)}

The possibility that the atom may decay by emitting a photon into a mode other than $a$ or $b$ can be approximately treated by making the replacement $\delta \to - i \gamma + \delta$ in Eqs.~(\ref{e3}) and (\ref{e4}) (note that $\gamma$ here is an \emph{amplitude} decay rate).  This ``pure-state approximation,'' corresponding to evolution with a non-Hermitian Hamiltonian (Eq.~(\ref{e1}) with $\delta \to - i \gamma + \delta$), ignores the fact that the atom must return to the ground state after a spontaneous emission event.  It is, nevertheless, often used in quantum optics to treat weakly-driven systems, if one is only concerned with effects of first-order in the driving field, since changes to the ground state population are of second order in the driving, but for our strongly-coupled system it would not do (beyond, perhaps, providing an order-of-magnitude estimate): a proper calculation of the gate fidelity for our system, in the presence of spontaneous emission losses, requires a full density-matrix treatment (see, e.g., \cite{scully,carmichael}). 

Nevertheless, as shown in the Appendix, we have found that this full treatment simplifies to some extent for our system, because it is not driven externally, and hence the Hamiltonian preserves the excitation number, while the (irreversible) spontaneous decay only couples manifolds of states with different excitation numbers in the downward direction (i.e., from two excitations to one to zero).  As a result of this, one can just use the pure-state approximation (with the replacement $\delta \to - i \gamma + \delta$) to calculate the evolution in the two-excitation manifold, and then use the terms obtained in that way as source terms for the evolution in the lower manifolds.  The final result for the gate fidelity ends up including the same terms shown in Eq.~(\ref{e12}), only now calculated from the non-Hermitian Hamiltonian evolution, plus a few additional terms:
\begin{align}
{\cal F} &= \text{[Eq.~(13) with $\delta \to - i \gamma + \delta$]} + \frac{1}{10}\rho^{(1)}_{00g,00g}\cr
&\quad + \frac{1}{20}\rho^{(2)}_{00g,00g}  + \frac{1}{10}\rho^{(2)}_{00e,00e}+\frac{1}{10}\rho^{(2)}_{01g,01g} 
\label{e16}
\end{align}
Here, $\rho^{(1)}_{00g,00g}$ (probability to be in the ground state with zero photons, when starting from the ground state with one photon) is calculated from the single-photon results as
\begin{equation}
\rho^{(1)}_{00g,00g} = 2\gamma \int_0^T \left|C_e^{(1)}(t)\right|^2\, dt
\label{e17}
\end{equation}
and the other terms correspond to the two-photon case (equations of motion for them are given in the Appendix).  Note that the symmetry of the system has been used throughout; in particular, $\frac{1}{10}\rho^{(2)}_{10g,10g} = \frac{1}{10}\rho^{(2)}_{01g,01g}$, where the first pair of subscripts refer to the $a$ and $b$ photons, respectively.

Figure 3 shows the gate fidelity as a function of $\gamma$ calculated from the density matrix, optimized at every point over $T$ and $\delta$, with $0\le gT \le 20$ and $0\le \delta/g \le 2$ (the parameter region covered in Figure 2).  Specifically, in the range $0\le \gamma/g \le 0.005$, we have taken $gT = 18.01$, $\delta =1.388g$; between $\gamma = 0.005g$ and $\gamma =0.015g$, we take $gT = 8.76$, $\delta =0.7g$; between $\gamma = 0.015g$ and $\gamma =0.07g$, $gT = 6.473$, $\delta =0$; and between $\gamma = 0.07g$ and $\gamma =0.155g$, $gT = 2.695$, $\delta =0$.  These choices of $T$ and $\delta$ roughly correspond to the different maxima shown, for $\gamma=0$, in Fig.~2; note how as $\gamma$ increases, a shorter time evolution is favored, as well as (eventually) the choice $\delta = 0$.


\medskip

\begin{figure}[h]
    \includegraphics[width=8cm]{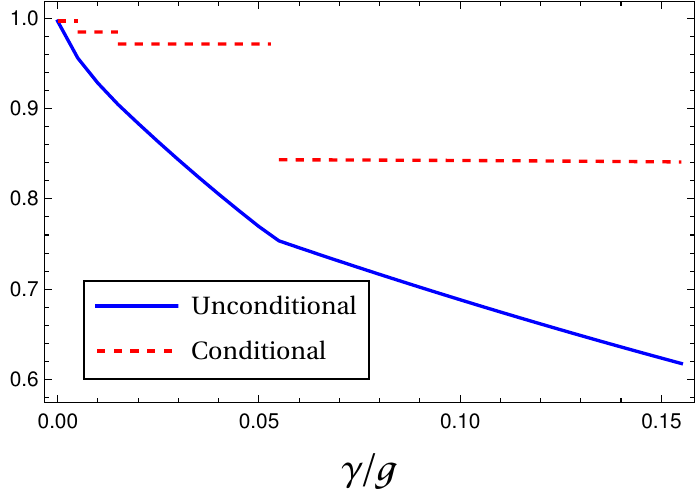}%
    {\caption{Solid line: The unconditional gate fidelity (\ref{e16}) as a function of $\gamma/g$, for optimal values of $gT$ and $\delta/g$. Dashed lines: the conditional gate fidelity for the same parameters. }}
\end{figure} 

Besides the ``unconditional'' fidelity just discussed, one may be interested in the \emph{conditional} fidelity, that is, the (gate) fidelity that one would obtain in a run of the experiment in which \emph{no} photons were lost to spontaneous emission.  This can be calculated easily, by ignoring the additional terms in Eq.~(\ref{e16}) and renormalizing the pure-state wavefunction (with $\delta \to - i \gamma + \delta$) before calculating (\ref{e12}).  The result is shown as the dashed line in Figure 3, for the same values of $T$ and $\delta$ as the corresponding conditional fidelity.  The stepwise decreases seen there are due to the fact that the successive choices of $T$ and $\delta$ made as $\gamma$ increases become optimal because they reduce the probability of a spontaneous emission relative to the previous choice, but they do lead to a smaller gate fidelity when spontaneous emission does not happen at all.

Figure 3 clearly shows that one needs to have a very small ratio of $\gamma$ to $g$ in order to have a substantial unconditional fidelity in this setup.  This was to be expected, since, as we showed in the previous subsection, getting a high fidelity in this system requires a relatively large value of $gT$, and the probability of a spontaneous emission event over the time $T$ (causing the loss of a photon, and hence an unavoidable decrease in fidelity) will scale as $2\gamma T = 2gT (\gamma/g)$.  For this reason, in the next couple of sections we will discuss schemes that have been proposed to either enhance $g$ or reduce $\gamma$, always in the context of a single-mode treatment of the two quantized fields.  

\section{Multiple V-type atoms}
It is well known that, for some atom-field interaction processes, having a large number of atoms $N$ at a given field location (in a volume small compared to the wavelength) leads to an effective enhancement of the atom-field coupling $g$ by a factor of $\sqrt N$.  However, while this is true for linear processes, and even for some nonlinear processes when the density of photons is sufficiently large, it does not work for single-photon nonlinear processes like the one considered here.  

The key fact that needs to be appreciated is that, in this scheme, in order for one photon to affect the other they both need to be interacting with the \emph{same} atom.  The essence of the V-atom nonlinearity is that an individual atom cannot absorb, say, an $a$ photon if it has absorbed a $b$ photon. Introducing more atoms would indeed make it possible for any of them to interact with either photon (thus increasing the single-photon coupling), but it would also make it much more likely for the two photons to interact with \emph{different} atoms, in which case the joint interaction presented in Section II.A would just not happen.  We should then expect the effective nonlinearity (specifically, the nonlinear phase shift $\phi_2-2\phi_1$) to actually \emph{go down} as $N$ increases.  

This can indeed be shown to be the case, formally, as follows.  Let the Hamiltonian for the $N$-atom, two-photon system be
\begin{align}
H= &\hbar \delta \sum_{i=1}^N\ket{e_a}_i\bra{e_a} +  \hbar \delta \sum_{i=1}^N\ket{e_b}_i\bra{e_b} \cr
&+ \hbar g\sum_{i=1}^N \Bigl(\ket{e_a}_i\bra{g} a + \ket{e_b}_i\bra{g} b + H.c. \Bigr)
\label{N1}
\end{align}
where the bras and kets shown act only on the space of the $i$-th atom.  This is a straightforward generalization of the single-atom Hamiltonian, under the assumption that all the atoms are close enough (well within a wavelength) to see the same field and hence the same coupling constant $g$ \cite{usc}; we further assume, as in Section II, that both transitions have identical strengths and detunings.  The state vector of the system when only one photon (say, $a$) is initially present and the atoms all start in the collective ground state $\ket{g_\text{all}}$ can be written as
\begin{equation}
\ket{\Psi(t)}^{(1)}  = C_g^{(1)}(t) \ket{g_\text{all}} \ket{1}_a + C_e^{(1)}(t) \ket{\psi_a} \ket{0}_a
\label{N3}
\end{equation}
Here
\begin{equation}
\ket{\psi_a} = \frac{1}{\sqrt{N}}\left( \sum_{i=1}^N  \ket{e_a}_i\bra{g}\right) \ket{g_\text{all}}
\label{psia}
\end{equation}
denotes a normalized, completely symmetric state in which one of the atoms is in the excited state $\ket{e_a}$, and all the others are in the ground state.  The state $\ket{\psi_b}$ is defined analogously.  When both photons are initially present, the evolution of the system is 
\begin{align}
\ket{\Psi(t)}^{(2)}  = &C_g^{(2)}(t) \ket{g_\text{all}} \ket{11}_{ab} + C_e^{(2)}(t) \ket{\psi_a} \ket{01}_{ab} \cr
&+ C_e^{(2)}(t) \ket{\psi_b} \ket{10}_{ab} +C_{ee}^{(2)}(t)\ket{\psi_{ab}} \ket{00}_{ab} \cr
\label{N5}
\end{align}
where
\begin{align}
\ket{\psi_{ab}}  
&=\frac{1}{\sqrt{N-1}} \left(\sum_{i=1}^N  \ket{e_a}_i\bra{g}\right) \ket{\psi_b} \cr 
&=\frac{1}{\sqrt{N-1}} \left(\sum_{i=1}^N  \ket{e_b}_i\bra{g}\right) \ket{\psi_a}
\label{psiab}
\end{align}
is again a symmetric, normalized state in which one atom is in state $\ket{e_a}$, another one in state  $\ket{e_b}$, and the rest in the ground state.  From the basic nature of $\ket{\psi_a}, \ket{\psi_b}$ and $\ket{\psi_{ab}}$, and taking into account their normalization, it is clear that the following results hold:
\begin{align}
\left(\sum_{i=1}^N \ket{e_a}_i\bra{e_a}\right) \ket{\psi_a} &= \ket{\psi_a} \cr
\left(\sum_{i=1}^N \ket{e_a}_i\bra{e_a}\right) \ket{\psi_{ab}} &= \ket{\psi_{ab}} \cr
\left(\sum_{i=1}^N \ket{g}_i\bra{e_a}\right) \ket{\psi_a} &= \sqrt N \ket{g_\text{all}}\cr
\left(\sum_{i=1}^N \ket{g}_i\bra{e_a}\right)\ket{\psi_{ab}} &= \sqrt{N-1} \ket{\psi_b}
\end{align}
so the Schr\"odinger equation, with the Hamiltonian (\ref{N1}), yields
\begin{align}
\dot C_{e}^{(1)} &= - i g\sqrt N\, C_g^{(1)} - i \delta  C_{e}^{(1)}\cr
\dot C_{g}^{(1)} &= -  i g \sqrt N\, C_{e}^{(1)}\cr
\label{M22}
\end{align}
for the single-photon case, and
\begin{align}
\dot C_{ee}^{(2)} &= -2i\delta C_{ee}^{(2)} -2ig\sqrt{N-1}\, C_e^{(2)} \cr
\dot C_{e}^{(2)} &= -i\delta C_{e}^{(2)} -ig\sqrt{N-1}\, C_{ee}^{(2)} - ig\sqrt{N}\,C_g^{(2)} \cr
\dot C_{g}^{(2)} &=  -2ig\sqrt{N}\, C_e^{(2)} 
\label{M18}
\end{align}
for the two-photon case.

In the limit where $N$ is large, so that one can approximate $\sqrt{N-1} \simeq \sqrt N$, it is easy to see that the solutions of Eqs.~(\ref{M22}) and (\ref{M18}) satisfy
\begin{equation}
C_{g}^{(2)} = {C_{g}^{(1)}}^2, \qquad C_{e}^{(2)} = {C_{g}^{(1)}}{C_{e}^{(1)}}, \qquad C_{ee}^{(2)} =  {C_{e}^{(1)}}^2
\label{M19}
\end{equation}
which means the $N$-atom two-photon system reduces, formally, to two independent $N$-atom, one-photon systems.  In particular, we see from the first of Eqs.~(\ref{M19}) that one will always have  $\phi_2 = 2 \phi_1$, and hence \emph{no} CPHASE at all in this limit.

We do note that this result depends on the approximation $\sqrt{N-1} \simeq \sqrt N$, which even for large $N$ will cease to be valid for long enough times, such that $gt/\sqrt{N} \sim 1$; however, the whole point of bringing in $N$ atoms was to increase the effective coupling so one could have a substantial effect for \emph{shorter} times, i.e., times such that $g\sqrt N t \sim 1$ while $\gamma t \ll 1$.  If one has to wait for times $t\sim \sqrt N/g$, then the requirement $\gamma t \ll 1$ becomes even harder to satisfy than in the single-atom problem.  

Although this clearly shows that a large number of atoms is undesirable, one may still wonder about what happens for a small number of atoms. For $\delta\ne 0$, solving the system (\ref{M18}) analytically requires solving a cubic equation that is, in general, much too unwieldy to be useful, but for $\delta=0$ the problem simplifies substantially and the solution for $C_g^{(2)}$ is
\begin{equation}
C_g^{(2)}(t) = \frac{N-1+N\cos(\sqrt{4N-2}\,gt)}{2N-1}
\end{equation}
It is clear that this can never be equal to $-1$, unless $N=1$; in fact, the largest (in magnitude) negative value it can take is $-1/(2N-1)$, which is only equal to $-1/3$ for $N=2$ and decreases monotonically in magnitude with $N$.  This means that, at least for $\delta =0$, the $N>1$ atom system, unlike the $N=1$ case, cannot get arbitrarily close to unit fidelity.
  
\medskip

\begin{figure}[h]
    \includegraphics[width=8cm]{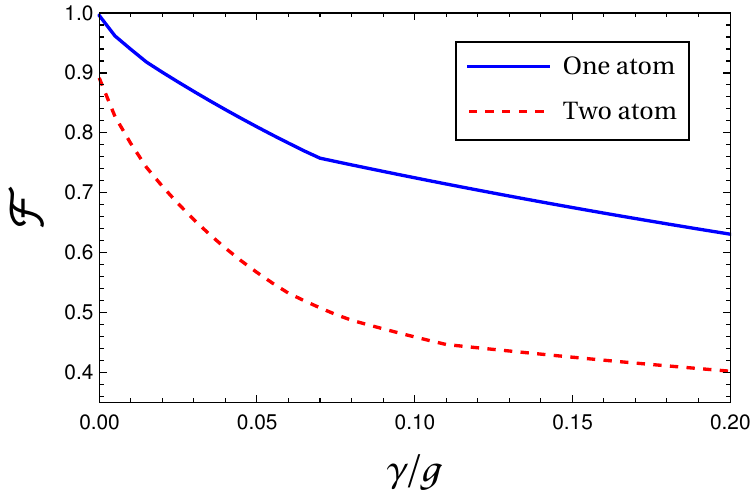}%
    {\caption{Solid line: The unconditional gate fidelity as a function of $\gamma/g$, for optimal values of $gT$ and $\delta/g$, for a single atom (Eq.~(\ref{e16})). Dashed line: the unconditional gate fidelity (see Appendix for the explicit formula), for two atoms, also optimized with respect to $gT$ and $\delta/g$. }}
\end{figure} 

Figure 4 shows the results of a full numerical study of the case $N=2$, including the effects of detuning and spontaneous emission losses, as a dashed line, compared to the equivalent result for a single atom (solid line).  
For each point, ${\cal F}$ has been optimized with respect to both $gT$ and $\delta/g$, over the same space of parameters shown in Figure 2, and the density matrix treatment has been used for both calculations (details  can be found in Appendix A.2).  It is apparent that adding even just one atom to the system substantially degrades its performance as a CPHASE gate, especially as the spontaneous emission losses increase.

\section{The 5-level scheme with two classical fields}
It has long been known that electromagnetically-induced transparency (EIT) can increase the effective optical nonlinearity of an atomic gas while at the same time decreasing its absorption, i.e., making it more transparent \cite{schmidt}, and in fact a proposal to use this  ``giant Kerr effect'' for quantum logic was put forth by Lukin and Imamoglu \cite{lukin}.  This idea motivated the authors of \cite{ottaviani,rebic} to consider the potential for a CPHASE gate of the 5-level, ``M''-configuration scheme illustrated in figure 5, where the two auxiliary levels $\ket{g_a}$ and $\ket{g_b}$ are coupled by external, classical fields (with Rabi frequencies $\Omega$) to the excited states $\ket{e_a}$ and $\ket{e_b}$, to provide EIT in the $\ket g \to \ket{e_a}$ and $\ket g \to \ket{e_b}$ transitions. The purpose of this section is to explore this system fully, in the single-atom regime (it is easy to verify that the argument against multiple atoms presented in the previous section  applies to this scheme as well \cite{rebic2}).

\begin{figure}[h]
    \includegraphics[width=8cm]{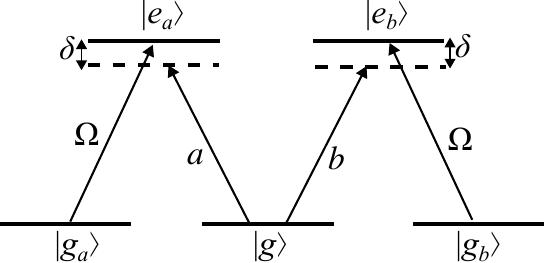}%
    {\caption{The 5-level system.}}
\end{figure} 

It may be worth pointing out, at the outset, that it is not immediately obvious how EIT would actually help here.  The standard derivation of EIT involves a perturbative treatment of the probe field (in this case, the single-photon fields $a$ and $b$) \cite{gea,hau}.  This implicitly assumes that the interaction of each probe photon with each atom is relatively weak, and hence both the absorption and the phase shift result from the cumulative effect of many atoms interacting with (essentially) a classical probe field. This is the complete opposite of the situation considered here, where each photon needs to be coupled as strongly as possible to a single atom.  Indeed, the results we show below are not really very EIT-like, and we believe it is best to just think of the auxiliary fields and levels as a way to introduce an additional parameter in the system---formally, the Rabi frequency $\Omega$---that makes it possible, to some extent, to satisfy the conditions (\ref{e8}) and (\ref{e9}) somewhat better than the three-level system in the presence of losses.

Figures 6 and 7 illustrate these points.  Figure 6 shows the gate fidelity for the 5-level system, as a function of $\gamma$, for different values of the auxiliary fields $\Omega$.  We find that, for small $\gamma$, the 5-level system always performs worse than the 3-level system. For $\gamma T$ larger than about $0.07$ in the figure, however, it is possible to find a value of $\Omega$ that brings the 5-level fidelity somewhat above the 3-level curve (which here corresponds to $\Omega =0$). Nevertheless, it appears that as $\Omega$ increases past some optimal value, the improvement over the $\Omega=0$ case disappears, or is confined to larger and larger values of $\gamma$, where the fidelity is already quite low.  Figure 7, which shows the gate fidelity as a function of $\Omega$, for different values of $\gamma$, confirms this and also suggests the existence of an optimum value, or range of values, of $\Omega$.  

\begin{figure}[h]
    \includegraphics[width=8cm]{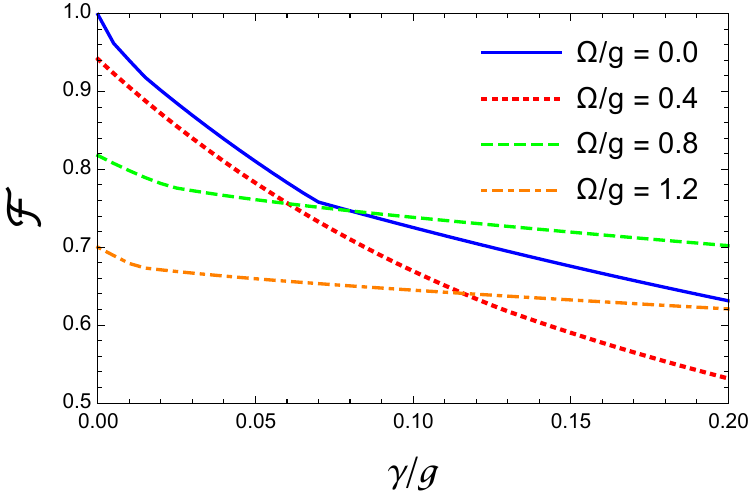}%
    {\caption{Gate fidelity for the 5-level system, as a function of $\gamma$, for different values of $\Omega$ (optimized for $T$ and $\delta$ in the intervals $0\le gT\le 30$ and $0\le \delta\le 10$).}}
\end{figure} 

\begin{figure}[h]
    \includegraphics[width=8cm]{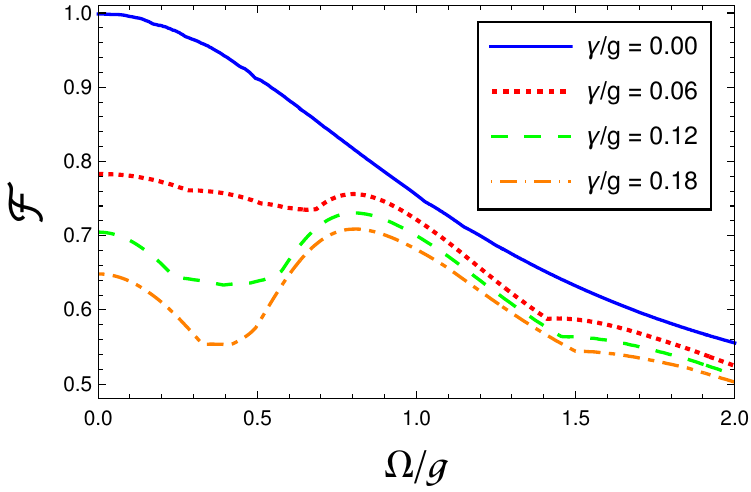}%
    {\caption{Gate fidelity for the 5-level system, as a function of $\Omega$, for different values of $\gamma$ (optimized for $T$ and $\delta$ in the intervals $0\le gT\le 30$ and $0\le \delta\le 10$).}}
\end{figure} 

As it turns out, however, Figures 6 and 7 do not tell the whole story. Each point in the graphs has been optimized with respect to $T$ and $\delta$ in the intervals $0\le gT\le 30$ and $0\le \delta/g\le 10$, but many of the values shown correspond, in fact, to either $gT=30$ or $\delta/g = 10$, meaning that larger values are possible in principle if either $T$ or $\delta$ are increased.  We find, in fact, that the optimal values, for given $\Omega$ and $\gamma$, are found by increasing both $gT$ and $\delta/g$ together, keeping the ratio $g^2T/\delta$ constant. In practice, of course, we would expect the maximum value of $T$ to be limited by some practical considerations (for instance, if the interaction takes place in a cavity, by the decay time of the field in the cavity, which we have not considered here at all), and similarly $\delta$ may be limited by the possibility of coupling to nearby atomic levels, so figures 6 and 7 are probably a fair representation of what one might qualitatively expect in a realizable experimental setting.  Nevertheless, a full study of the asymptotic behavior of the 5-level system, for large $T$ and $\delta$, is possible and not exempt of interest, so we will devote the rest of this section to it.

We begin with the equations of motion for the wavefunction amplitudes in the ``quasi-pure state'' approximation.  When both photons are initially present, we have
\begin{align}
\dot{C}_{ea}^{(2)} &=  - (\gamma + i \delta) {C}_{ea}^{(2)} - i g {C}_{g}^{(2)} -i\Omega {C}_{ga}^{(2)}\cr
\dot{C}_{g}^{(2)}  &= -i g  {C}_{ea}^{(2)} - i g {C}_{eb}^{(2)} \cr
\dot{C}_{eb}^{(2)} &= -(\gamma + i \delta) {C}_{eb}^{(2)} - i g {C}_{g}^{(2)} -i\Omega {C}_{gb}^{(2)} \cr
\dot{C}_{ga}^{(2)} &= -i \Omega {C}_{ea}^{(2)} \cr
\dot{C}_{gb}^{(2)} &= -i \Omega {C}_{eb}^{(2)} 
\label{e26} 
\end{align}
and when only one is present (say, the $a$ photon) we have
\begin{align}
\dot{C}_{ea}^{(1)} &=  -(\gamma + i \delta) {C}_{ea}^{(1)} - i g {C}_{g}^{(1)} -i\Omega {C}_{ga}^{(2)}\cr
\dot{C}_{g}^{(1)}  &= -i g  {C}_{ea}^{(2)} \cr
\dot{C}_{ga}^{(1)} &= -i \Omega {C}_{ea}^{(2)} 
\label{e27} 
\end{align}
Both (\ref{e26}) and (\ref{e27}) can be solved with the initial condition $C_g(0) =1$, with the results, for the ground state amplitude at the time $t$,
\begin{align}
{C}_{g}^{(1)} &= \frac{\Omega^2}{g^2+\Omega^2} + \frac{g^2}{g^2+\Omega^2} e^{-\frac 1 2 t(\gamma+i\delta)} \times\cr
&\quad\frac 1 2\left [ e^{-\mu_1 t/2}\left(1-\frac{\gamma+i\delta}{\mu_1}\right) +  e^{\mu_1 t/2}\left(1+\frac{\gamma+i\delta}{\mu_1}\right)\right],\cr
\mu_1 &\equiv \sqrt{(\gamma+i\delta)^2-4\Omega^2-4g^2}
\label{e29}
\end{align}
and
\begin{align}
{C}_{g}^{(2)} &= \frac{\Omega^2}{2 g^2+\Omega^2} + \frac{g^2}{2 g^2+\Omega^2} e^{-\frac 1 2 t(\gamma+i\delta)} \times\cr
&\quad\left [ e^{-\mu_2 t/2}\left(1-\frac{\gamma+i\delta}{\mu_2}\right) +  e^{\mu_2 t/2}\left(1+\frac{\gamma+i\delta}{\mu_2}\right)\right],\cr
\mu_2 &\equiv \sqrt{(\gamma+i\delta)^2-4\Omega^2-8g^2}
\label{e30}
\end{align}
These results immediately show that, if $\Omega$ is allowed to become very large, one will simply have ${C}_{g}^{(1)} = {C}_{g}^{(2)} = 1$, and the desired nonlinear phase shift will vanish.  This makes sense physically: a very large $\Omega$ produces a Stark shift that takes the atom out of resonance with the $a$ and $b$ photons, in such a way that the interaction effectively vanishes.  

The interesting thing, however, is that it is formally possible to take the photons very far from resonance in another way, by making $\delta$ large---so large that, in effect, the probability of a spontaneous emission event becomes negligible---and yet, as long as $\Omega$ remains finite, one can still approximately achieve the desired phase shift, albeit for very long times. 

The precise result follows in a very straightforward way from Eqs.~(\ref{e29}) and (\ref{e30}).  Note that for $\delta \gg \gamma, g, \Omega$ one has
\begin{align}
\mu_1 &= \gamma + i\delta +\frac{2i}{\delta}(g^2+\Omega^2) + O\left(\frac{1}{\delta^2}\right)\cr
\mu_2 &= \gamma + i\delta +\frac{2i}{\delta}(2g^2+\Omega^2) + O\left(\frac{1}{\delta^2}\right)
\label{e31}
\end{align}
and consequently we have the asymptotic forms
\begin{align}
{C}_{g}^{(1)} &\simeq \frac{\Omega^2}{g^2+\Omega^2} + \frac{g^2}{g^2+\Omega^2} e^{it (g^2+\Omega^2)/\delta}\cr
{C}_{g}^{(2)} &\simeq \frac{\Omega^2}{2g^2+\Omega^2} + \frac{2g^2}{2g^2+\Omega^2} e^{it (2g^2+\Omega^2)/\delta}
\label{e32}
\end{align}
One can see now how the conditions (\ref{e8}) and (\ref{e9}) can be approximately satisfied. To begin with $\Omega/g$ should be sufficiently small for the second term on the right-hand side of Eqs.~(\ref{e32}) to dominate over the first; then, to get Eq.~(\ref{e9}), it would suffice to have
\begin{equation}
\frac t \delta(2 g^2+\Omega^2) -2 \frac t \delta(2 g^2+\Omega^2) = -\frac t \delta\Omega^2 = -n\pi
\label{e33}
\end{equation}
with $n$ odd.  Note that with small $\Omega/g$ this can only be satisfied for very large values of $gt$; $\delta/g$ itself is required to be very large in order for the approximation (\ref{e31}) to be valid, and we require, additionally, $\delta \gg \gamma,\Omega$. The value $\delta=10g$ in the optimized fidelity plots of Figs.~6 and 7 is, in fact, sufficiently large for (\ref{e31}) to be approximately valid; the optimum value of $\Omega$ seen then in the graphs, about $0.8g$, is the best compromise between trying to keep $\Omega$ small enough for $|C_g| \sim 1$ in Eqs.~(\ref{e32}), and large enough for (\ref{e33}) to be approximately valid, given the constraint $gT\le 30$.

\section{A two-level atom scheme.}
The three- (or five-) level scheme considered so far is suitable for a CPHASE gate when a ``single-rail'' encoding is used (i.e., the logical $0$ and $1$ states correspond to single photon states with orthogonal polarizations \cite{note4}).  However, by making use of a setup such as the one shown in \cite{nysteen}, one could apply these ideas to a dual-rail encoding, the idea being that an initial state $\ket{\epsilon_1,\epsilon_2}$, with $\epsilon_1,\epsilon_2 \in \{0,1\}$, becomes a cavity field state with $\epsilon_1+\epsilon_2$ photons.  Then a single two-level atom can be used to produce the desired phase shift between the single-photon and two-photon states.  

With a single two-level atom in the cavity, the results for $C_g^{(1)}(t)$ and $C_g^{(2)}(t)$ turn out to be identical to those given by Eqs.~(\ref{e5})--(\ref{e7}). The expression for the gate fidelity is also identical to Eq.~(\ref{e12}), except for the absence of the terms involving the excited state amplitudes.  This is because, with a dual-rail encoding, the number of physical photons involved in a two-qubit operation is always two, regardless of the initial logical state.  Hence, if the atom is left in an excited state at the end of the interaction time, the final field state will necessarily be orthogonal to the ideal one, since it will have one photon less.  This means that the gate fidelity will always be slightly smaller than for the single-rail, three-level scheme, although it, too, can in principle be made arbitrarily large for sufficiently large times.

As was the case for the system considered in section III, here also adding more atoms has a detrimental effect, and eventually causes the nonlinearity to vanish.  If we write the state of the $N$-atom, 1-photon system in the form
\begin{equation}
\ket{\Psi\left(t\right)}^{(1)} = C_{e}^{(1)} |0\rangle \ket{\psi_e} + C_{g}^{(1)} |1\rangle \ket{g_\text{all}} 
\label{e34}
\end{equation}
where $\ket{\psi_e}$ is defined in a form analogous to $\ket{\psi_a}$ in Eq.~(\ref{psia}), the equations of motion for 
$C_{e}^{(1)}$ and $C_{g}^{(1)}$ are identical to Eqs.~(\ref{M22}).  On the other hand, for the $N$-atom, 2-photon case, the overall state must be written as
\begin{equation}
\ket{\Psi\left(t\right)}^{(2)} = C_{ee}^{(2)} |0\rangle \ket{\psi_{ee}} +C_{e}^{(2)} |1\rangle \ket{\psi_{e}}+ C_{g}^{(2)} |2\rangle \ket{g_\text{all}} 
\label{e35}
\end{equation}
where now instead of Eq.~(\ref{psiab}) we must define
\begin{equation}
\ket{\psi_{ee}} = {\frac{1}{\sqrt{2(N-1)}}} \left( \sum_{i=1}^N \ket{e}_i\bra{g} \right) \ket{\psi_e}
\label{e36}
\end{equation}
and the equations of motion read
\begin{align}
\dot C_{ee}^{(2)} &= -2i\delta C_{ee}^{(2)} -ig\sqrt{2(N-1)}\, C_e^{(2)} \cr
\dot C_{e}^{(2)} &= -i\delta C_{e}^{(2)} -ig\sqrt{2(N-1)}\, C_{ee}^{(2)} - ig\sqrt{2N}\,C_g^{(2)} \cr
\dot C_{g}^{(2)} &=  -ig\sqrt{2N}\, C_e^{(2)} 
\label{e37}
\end{align}
Now it is easy to see that, in the limit $N\gg 1$, the solution to the system (\ref{e37}) can be written in terms of the solution to the system (\ref{M22}) as
\begin{equation}
C_{g}^{(2)} = {C_{g}^{(1)}}^2, \qquad C_{e}^{(2)} = \sqrt 2{C_{g}^{(1)}}{C_{e}^{(1)}}, \qquad C_{ee}^{(2)} =  {C_{e}^{(1)}}^2
\label{e38}
\end{equation}
and therefore, as before, $\phi_2 = 2 \phi_1$.

As mentioned in the Introduction, the authors of \cite{heuck2} did briefly consider a scheme in which the nonlinear phase  shift would result from the interaction with a two-level atom, rather than a second- or third-order nonlinearity.  In their scheme, a control field and a nonlinear medium would be used to bring the frequency of the cavity field, initially off-resonance with the atom, into resonance, and off again.  They found numerically a shape for a control pulse that achieved unit fidelity, under lossless conditions.  Our scheme here may be regarded as a variation on theirs, where we assume the atom-field interaction is turned on and off abruptly, as opposed to gradually.

\section{Conclusions}
The goal of this research was to consider---and, where necessary, clarify---the potential for quantum logic of systems of the Jaynes-Cummings type, i.e., single or multiple atoms interacting with single temporal modes, containing one or two photons, over a finite time.  Our two main results are: that (within the scope of such models), gate fidelities arbitrarily close to 1 can be achieved for some of these systems if losses (including cavity losses) are neglected; and that using more than one atom is suboptimal, and a very large number of atoms actually causes the useful phase shift to go to zero.

Our analysis has neglected entirely the cavity losses, i.e., it has assumed essentially a cavity of ``infinite'' finesse at the operating frequency \cite{raymer}.  We expect that cavity losses will degrade the system's performance in a way similar to spontaneous emission losses (see also the loss analysis in \cite{heuck1}).  Alternatively, we could say that the fidelities we have calculated are all actually conditioned on no photons being lost through cavity losses.

Optical cavities with low values of $\gamma/g$, and also (to a lesser extent) $\kappa/g$, have been reported: for example, $\gamma/g = 0.45$ and $\kappa/g = 0.37$ for the conventional optical cavity in \cite{reiserer}, and $\gamma/g = 0.014$ and $\kappa/g = 0.25$ for the ``optical cavity on a chip'' of \cite{colombe}. In principle, if the latter value of $\kappa/g$ could be reduced by an order of magnitude, gate fidelities very close to 1 should be possible with our setup. We note that in a relatively recent review article Chang et al. \cite{kimble}  have projected that nanocavities could in the future achieve single-atom cooperativities ($g^2/\kappa\gamma$) of the order of $10^3$ and even $10^4$ (see Fig.~6 of \cite{kimble}), although it is not clear how this improvement of up to two orders of magnitude over recent values would be split between $g/\kappa$ and $g/\gamma$.  

Although the above numbers appear hopeful, the real question is how large an atom-cavity coupling could be achieved in the dynamically-coupled setups envisioned in \cite{heuck1,heuck2,raymer}.  We are unable to answer this, but note that a single two-level atom was already considered as a potential candidate for a nonlinear medium in \cite{heuck2}, and also that the nanocavities envisioned in \cite{kimble} would appear to be consistent, in this regard at least, with the small mode volumes that appear to be required in the analysis in \cite{heuck1,heuck2}. 

Another relevant question is that, since any attempt to realize this system in practice would require making use of optical nonlinearities to load and unload the cavities, would there be anything to be gained by using an atom for the nonlinear phase shift, instead of the nonlinear materials themselves (as suggested in \cite{heuck1})?  An immediate answer is that the scheme in \cite{heuck1} requires, for the phase shift, very large optical nonlinearities at the single-photon level, whereas the loading and unloading of the cavity can in principle be achieved with reasonable nonlinearities, as long as the auxiliary field is strong enough.  Moreover, it is not immediately apparent why the large single-photon nonlinearities assumed in \cite{heuck1} would not suffer from the phase noise problem originally pointed out in \cite{shapiro,shapiro2}, at least for the $\chi^{(3)}$ case.  

The situation is different with regard to the two-level atom scheme presented in Section VI.C of \cite{heuck2}.  This can be regarded as a variation of our scheme involving a gradual turning on and off of the interaction.  From a theory viewpoint, our approach is simpler in that it allows for a largely analytical treatment, and does not require numerically searching for an optimal control pulse. From an experimental viewpoint, a fair comparison of the two approaches would probably require a detailed analysis based on specific parameters.  Such an analysis, however, is beyond the scope of the present paper.

\appendix

\section{density-matrix gate fidelity calculations}

\subsection{Single V-type atom}
The density matrix equation of motion for this system is 
\begin{equation}
\dot \rho = -\frac{i}{\hbar}[H,\rho] -\gamma\sum_{e=e_a,e_b}\bigl(\rho\ket e \bra e + \ket e \bra e \rho - 2 \ket g\bra e \rho \ket e \bra g \bigr)
\label{a1}
\end{equation}
In this expression, the first two terms under the summation sign give the decay of the excited states, and their action on those states (and/or on the corresponding density matrix elements) can be completely accounted for by a wavefunction treatment with the non-Hermitian Hamiltonian resulting from the substitution $\delta \to \delta-i\gamma$ in Eq.~(\ref{e1}) (the ``pure-state approximation'' mentioned in Section II.B).  The last term in (\ref{a1}), on the other hand, is a ``source'' term that repopulates the ground state as a result of the decay of an excited state.  It only acts on, and only produces,  diagonal components (in the atomic basis) of the density operator.  It cannot be handled by pure-state (Hamiltonian) methods, except when the master equation is unraveled along ``quantum trajectories,'' in which case its effect is accounted for by the random ``jumps'' that are eventually averaged over in that formalism \cite{scully85}.  We will not use such an unraveling here; instead, we show below how the nature of our system allows us to simplify the solution of Eq.~(\ref{a1}). 

As mentioned in the main text, our system has the property that the Hamiltonian evolution (in which we include now all but the last term of (\ref{a1})) preserves the excitation number (number of photons  $+$ number of excited states).  The action of the last term, on the other hand, reduces the excitation number by one.  This allows for a substantial simplification of the calculation of the gate fidelity (\ref{e11}): the evolution of $\rho$ in a manifold with a given excitation number is not affected by the terms evolving in a lower manifold, and, in particular, the evolution in the manifold with the largest excitation number can be calculated using a pure state with the non-Hermitian Hamiltonian.  Diagonal terms in a manifold then act, through the last term in (\ref{a1}), as source terms for the density matrix evolution in the next lower manifold.  

To calculate ${\cal F}$ one needs to apply Eq.~(\ref{a1}) to the density matrix that evolves from the initial state $\rho(0) = \ket{\Psi(0)}\bra{\Psi(0)}$, with 
\begin{equation}
\ket{\Psi(0)} = (\alpha_{00}\ket{00} +  \alpha_{01}\ket{01} + \alpha_{10}\ket{10} +  \alpha_{11}\ket{11})\ket g
\label{a2}
\end{equation}
By the linearity of the master equation, we can consider separately the evolution of each of the 16 terms into which $\ket{\Psi(0)}\bra{\Psi(0)}$ splits.  The diagonal terms all have a well-defined excitation number, and can be calculated by the approach sketched above.  The off-diagonal terms can mix different manifolds, but a careful study of the equations of motion shows that their contribution to the gate fidelity can also be calculated by considering only the non-Hermitian Hamiltonian evolution.  

Consider, for example, the terms that evolve from $\alpha_{01}\alpha_{11}^\ast \ket{01}\bra{11}\otimes \ket g \bra g$.  Initially, the last term in (\ref{a1}) has no effect on this component of the density operator, but with time the Hamiltonian evolution can transform it into $\alpha_{01}\alpha_{11}^\ast \ket{00}\bra{10}\otimes \ket{e_b}\bra{e_b}$, which by the action of the last term in (\ref{a1}) can then evolve into  $\alpha_{01}\alpha_{11}^\ast \ket{00}\bra{10}\otimes \ket{g}\bra{g}$.  However, when the expectation value in $\ket{\Phi_\text{ideal}}$ is taken this term will select the coefficients $\alpha_{00}^\ast$ on the left and $\alpha_{10}$ on the right (since the number of photons has gone down by 1, on either side, relative to the initial state), and the average of $\alpha_{00}^\ast\alpha_{01}\alpha_{11}^\ast \alpha_{10}$ is zero.

In this way, we eventually obtain the result (\ref{e16}) of the main text, where the term $\rho^{(1)}_{01g,01g}$ is calculated as shown in Eq.~(\ref{e17}), and the other terms correspond to the evolution in the 1-excitation manifold driven by spontaneous decay from the upper (2-photon) manifold, in the way described above:
\begin{align}
\dot\rho^{(2)}_{01g,01g} &= 2\gamma \rho^{(2)}_{01e_a,01e_a}-ig\left(\rho^{(2)}_{00e_b,01g}-\rho^{(2)}_{01g,00e_b}\right) \cr
\dot\rho^{(2)}_{00e_b,01g} &= -(\gamma+i\delta)\rho_{00e_b,01g} -ig\left(\rho^{(2)}_{01g,01g}-\rho^{(2)}_{00e_b,00e_b}\right) \cr
\dot\rho^{(2)}_{01g,00e_b} &= -(\gamma-i\delta)\rho_{01g,00e_b} +ig\left(\rho^{(2)}_{01g,01g}-\rho^{(2)}_{00e_b,00e_b} \right) \cr
\dot\rho^{(2)}_{00e_b,00e_b} &= -2\gamma \rho^{(2)}_{00e_b,00e_b}+ig\left(\rho^{(2)}_{00e_b,01g}-\rho^{(2)}_{01g,00e_b}\right) 
\label{a3}
\end{align}
(where $\gamma \rho^{(2)}_{01e_a,01e_a} = |C_{e_a}^{(2)}(t)|^2$).  A similar set of equations describes the evolution on the ``$b$'' side, switching $e_a$ and $e_b$ and the corresponding photonic subscripts. Two of the additional terms in (\ref{e16}) are given directly by the diagonal elements in  (\ref{a3}), and the last remaining one is given by  
\begin{equation}
\dot\rho^{(2)}_{00g,00g} = 2\gamma \rho^{(2)}_{00e_b,00e_b} + 2\gamma \rho^{(2)}_{00e_a,00e_a}
\label{a6}
\end{equation}

For completeness, we show below also how we have calculated the nonzero averages of the products of coefficients $\alpha_{ij}$.  Let $x=|\alpha_{00}|^2$, $y=|\alpha_{01}|^2$, $z=|\alpha_{10}|^2$.  Then, for the state (\ref{a2}) to be normalized, we must have $|\alpha_{11}|^2=1-x-y-z$.  As this quantity has to be between 0 and 1, we find $z\le 1-x-y$, and again because $z\ge 0$ we find $y\le 1-x$.  So, to calculate our averages, we can use a probability distribution function which is constant and nonzero over the volume defined by $\{0\le x\le 1\;\&\& \; 0\le y \le 1-x\; \&\&\; 0\le z\le 1-x-y \}$ (note that in spite of the seemingly asymmetric way we have defined this volume, it is in fact symmetric in the three coordinates, a triangular pyramid).  We normalize this by requiring that the average of 1 be 1, that is,
\begin{equation}
1 =\frac{1}{\cal N} \int_0^1 dx\int_0^{1-x}dy \int_0^{1-x-y}dz = \frac{1}{6\cal N}
\end{equation}
So, with ${\cal N} = 1/6$, we can calculate the averages we want, as
\begin{equation}
\overline{|\alpha_{ij}|^4} = \overline{x^2} = 6\int_0^1 x^2\,dx\int_0^{1-x}dy \int_0^{1-x-y}dz = \frac{1}{10}
\end{equation}
and 
\begin{equation}
\overline{|\alpha_{ij}|^2|\alpha_{kl}|^2} = \overline{xy} = 6\int_0^1 x\,dx\int_0^{1-x}y\,dy \int_0^{1-x-y}dz = \frac{1}{20}
\end{equation}

\subsection{Two V-type atoms}

For the 2-atom case, it is best to use the basis introduced in Section III, where the atomic state with one excitation of the ``$a$'' type is the symmetric combination
\begin{equation}
\ket{\psi_a} = \frac{1}{\sqrt 2} \bigl(\ket{e_a,g} + \ket{g,e_a}\bigr)
\end{equation}
and similarly for $\ket{\psi_b}$, and the doubly excited state is also the symmetric combination
\begin{equation}
\ket{\psi_{ab}} = \frac{1}{\sqrt 2} \bigl(\ket{e_a,e_b} + \ket{e_b,e_a}\bigr)
\end{equation}
In Eqs.~(\ref{M22}) and (\ref{M18}), $C_e$ denotes the probability amplitude to find the system in either one of $\ket{\psi_a}$ or $\ket{\psi_b}$, and $C_{ee}$ that of finding it in $\ket{\psi_{ab}}$ 

We shall also assume that the two atoms decay to the same reservoir, which allows us to stay within the atomic space spanned by $\{\ket{gg},\ket{\psi_a},\ket{\psi_b},\ket{\psi_{ab}}\}$.  The corresponding master equation is
\begin{equation}
\dot \rho = -\frac{i}{\hbar}[H,\rho] -\gamma\sum_{l=a,b}\bigl(\rho J_l^\dagger J_l + J_l^\dagger J_l\rho - 2 J_l\rho J_l^\dagger\bigr)
\label{a13}
\end{equation}
(see Eq.~(6.131) of \cite{carmichael}), where the $J_l$ are collective atomic decay operators, $J_a = \ket{g}_1\bra{e_a} + \ket{g}_2\bra{e_b}$ and $J_b = \ket{g}_1\bra{e_b} + \ket{g}_2\bra{e_b}$ .  They can also be defined by their effect on the states of interest: $J_a\ket{gg} = J_a\ket{\psi_b} = 0$,  $J_a\ket{\psi_a} = \sqrt 2\ket{gg}$, $J_a\ket{\psi_{ab}} = \ket{\psi_b}$ (and similarly for $J_b$).

As was the case for the single atom, the action of the $J^\dagger J$ terms alone could be accounted for by pure-state evolution under a modified non-Hermitian Hamiltonian
\begin{equation}
H^\prime = H -2i\gamma\Bigl(\ket{\psi_a}\bra{\psi_a} + \ket{\psi_b}\bra{\psi_b} + \ket{\psi_{ab}}\bra{\psi_{ab}}\Bigr)
\end{equation}
which means all the excited-state amplitudes in Eqs.~(\ref{M22}) and (\ref{M18}) would decay at the rate $2\gamma$ (a collective enhancement of the decay rate, relative to the $N=1$ case, that parallels the enhanced coupling to the field).  This Hamiltonian evolution again preserves the excitation number, whereas the $J\rho J^\dagger$ term in (\ref{a13}) drops the system down to a manifold with one fewer excitation.  This means that, as was the case for the single atom, the expression for the gate fidelity can be written as a sum of terms arising from the non-unitary Hamiltonian evolution of the uppermost manifold, plus terms arising from the evolution of the lower manifolds that do require solving the master equation, in a reduced subspace, and with source terms derived from the higher-excitation manifolds.

The Hamiltonian evolution can be computed by starting with the state 
\begin{equation}
\ket{\Psi(0)} = (\alpha_{00}\ket{00} +  \alpha_{01}\ket{01} + \alpha_{10}\ket{10} +  \alpha_{11}\ket{11})\ket{gg}
\label{a11}
\end{equation}
and building the time-evolved state $\ket{\Psi(t)}$ through the substitutions
\begin{align}
\ket{01}\ket{gg} &\to C_g^{(1)}(t) \ket{01}\ket{gg} +  C_e^{(1)}(t) \ket{00}\ket{\psi_b} \cr
\ket{10}\ket{gg} &\to C_g^{(1)}(t) \ket{10}\ket{gg} +  C_e^{(1)}(t) \ket{00}\ket{\psi_a} \cr
\ket{11}\ket{gg} &\to C_g^{(2)}(t) \ket{11}\ket{gg} +  C_e^{(2)}(t) \ket{01}\ket{\psi_a} \cr 
&\quad +  C_e^{(2)}(t) \ket{10}\ket{\psi_b} +  C_{ee}^{(2)}(t) \ket{00}\ket{ee} 
\end{align}
where the $C^{(1)}$ and $C^{(2)}$ coefficients are the solutions to Eqs.~(\ref{M22}) and (\ref{M18}), respectively, with the additional decay terms, and starting from the ground state. The trace of $\ket{\Psi(t)}\bra{\Psi(t)}$ over the atoms produces a $\rho_f(t)$ that is formally identical to the single-atom result, except for the extra term $|\alpha_{11}|^2 |C_{ee}^{(2)}(t)|^2 \ket{00}\bra{00}$, so taking the expectation value $\bra{\Phi_\text{ideal}}\rho_f \ket{\Phi_\text{ideal}}$ and averaging over the $\alpha_{ij}$ produces an expression for the gate fidelity that is also formally identical to Eq.~(\ref{e12}), except for an additional term 
\begin{equation}
\frac{1}{20}|C_{ee}^{(2)}(T)|^2
\end{equation}

To this one must now add the effect of the evolution caused by decay to lower-excitation manifolds.  As in the single-atom case, and for the same reasons, these do not contribute to off-diagonal (in the photon basis) terms, that is, to terms that start out from $\ket{ij}\bra{kl}$ with $i\ne k, j\ne l$.  For single-photon, diagonal terms (starting from $\ket{01}\bra{01}$ or $\ket{10}\bra{10}$), the additional terms are again trivial:  the term $2\gamma |C_e^{(1)}|^2 J_a \ket{\psi_a}\bra{\psi_a}J^\dagger_a = 4\gamma |C_e^{(1)}|^2 \ket{gg}\bra{gg}$ populates the double ground state at a rate $4\gamma |C_e^{(1)}|^2$, so together the $a$ and $b$ terms contribute to ${\cal F}$ the amount
\begin{equation}
2\times\frac{1}{20} \rho^{(1)}_{00g,00g} = \frac{2\gamma}{5}\int_0^T|C_{e}^{(1)}(t)|^2 dt
\end{equation}
(note that in subscripts we use the single letter $g$ to refer to the double ground state $\ket{gg}$, to lighten the notation, and for consistency with Section III).

For the two-photon case, spontaneous decay produces source terms proportional to $|C_e^{(2)}|^2J_a \ket{\psi_a}\bra{\psi_a}J^\dagger_a$, $|C_e^{(2)}|^2J_b \ket{\psi_b}\bra{\psi_b}J^\dagger_b$, $|C_{ee}^{(2)}|^2J_a \ket{\psi_{ab}}\bra{\psi_{ab}}J^\dagger_a$, and $|C_{ee}^{(2)}|^2J_b \ket{\psi_{ab}}\bra{\psi_{ab}}J^\dagger_b$.  The corresponding 1-excitation evolution is governed by the equations
\begin{align}
\dot\rho^{(2)}_{01g,01g} &= 4\gamma |C_e^{(2)}|^2-ig\sqrt 2\left(\rho^{(2)}_{00\psi_b,01g}-\rho^{(2)}_{01g,00\psi_b}\right) \cr
\dot\rho^{(2)}_{00\psi_b,01g} &= -(2\gamma+i\delta)\rho_{00\psi_b,01g} -ig\sqrt 2\left(\rho^{(2)}_{01g,01g}-\rho^{(2)}_{00\psi_b,00\psi_b}\right) \cr
\dot\rho^{(2)}_{01g,00\psi_b} &= -(2\gamma-i\delta)\rho_{01g,00\psi_b} +ig\sqrt 2\left(\rho^{(2)}_{01g,01g}-\rho^{(2)}_{00\psi_b,00\psi_b} \right) \cr
\dot\rho^{(2)}_{00\psi_b,00\psi_b} &= -4\gamma \rho^{(2)}_{00\psi_b,00\psi_b} +ig\sqrt 2\left(\rho^{(2)}_{00\psi_b,01g}-\rho^{(2)}_{01g,00\psi_b}\right) \cr
&\quad + 2\gamma |C_{ee}^{(2)}|^2
\label{a16}
\end{align}
on the $\ket{\psi_b}$ side, and a similar set involving $\ket{\psi_a}$, leading finally to
\begin{equation}
\dot\rho^{(2)}_{00g,00g} = 4\gamma \rho^{(2)}_{00\psi_b,00\psi_b} + 4\gamma \rho^{(2)}_{00\psi_a,00\psi_a}
\label{a17}
\end{equation}
The final expression for the gate fidelity will then be
\begin{align}
{\cal F}^2 &= \text{[as in Eq.~(13)]} + \frac{1}{20}|C_{ee}^{(2)}(t)|^2+\frac{1}{10}\rho^{(1)}_{00g,00g}\cr
&\quad + \frac{1}{20}\rho^{(2)}_{00g,00g}  + \frac{1}{10}\rho^{(2)}_{00e,00e}+\frac{1}{10}\rho^{(2)}_{01g,01g} 
\label{e18}
\end{align}
where the subscript ``$e$'' in the next to last term could stand for either $\psi_a$ or $\psi_b$ equivalently.

\subsection{Single M-type atom}

Unlike in the previous two cases, this is an externally-driven system, which means that the Hamiltonian evolution does not preserve the excitation number:  from the state $\ket{00}\ket{g_a}$, for instance, the external field $\Omega$ can take the system to $\ket{00}\ket{e_a}$ (and from here, again by Hamiltonian evolution, to $\ket{10}\ket g$).  This means that the evolution of the uppermost manifold is no longer insulated from the lower ones, and one ends up having to solve the full master equation repeatedly, starting with different initial conditions, to calculate the gate fidelity.  

Specifically, the master equation for this case takes the form
\begin{align}
\dot \rho = &-\frac{i}{\hbar}[H,\rho] -\gamma\sum_{l=a,b}\Bigl[\rho\ket{e_l} \bra{e_l} + \ket{e_l} \bra{e_l} \rho \cr
& \qquad - \ket g\bra{e_l} \rho \ket{e_l} \bra g -\ket{g_l}\bra{e_l} \rho \ket{e_l} \bra{g_l} \Bigr]
\label{a18.1}
\end{align}
where, to make the comparison to the three-level system as favorable for the 5-level scheme as possible, we have kept the total amplitude decay rate of the excited states equal to $\gamma$, even though they now have two states to decay into.  

The final expression for the gate fidelity in this case takes the form
\begin{equation}
{\cal F} = \frac{1}{10}\left( 1 + A + B + C + D\right)
\label{a23}
\end{equation}
where, as always, the $1$ comes from the $\ket{00}\bra{00}$ term in $\ket{\Psi(0)}\bra{\Psi(0)}$, and
\begin{equation}
A = 2|C_g^{(1)}| -\frac{1}{|C_g^{(1)}|^2 } \text{Re}\left[\left({ C_g^{(1)}}^\ast\right)^2  C_g^{(2)}\right] + |C_g^{(1)}|^2
\end{equation}
are the only terms that can be obtained from the non-Hermitian Hamiltonian evolution, which in this case involves solving the system (\ref{e26}) and (\ref{e27}).  Here, the first term comes from the $\ket{00}\bra{01}$ and $\ket{00}\bra{10}$ terms, the second one comes from $\ket{00}\bra{11}$, and the last one comes from $\ket{01}\bra{10}$ (and Hermitian conjugates in every case).

All the other terms in (\ref{a23}) require solving the full master equation.  The terms in $B$ are
\begin{equation}
B=2\rho^{(1)}_{01g,01g} +\rho^{(1)}_{00e_b,00e_b}+\rho^{(1)}_{00g_b,00g_b}+\rho^{(1)}_{00g,00g}
\end{equation}
and they arise from $\ket{01}\bra{01}$ (or equivalently $\ket{10}\bra{10}$), that is, from solving the master equation with the initial condition $\rho_{01g,01g} =1$ (the $a$ and $b$ contributions are, of course, identical, and have been added already in the expression (\ref{a23})). 

The term in $C$ is
\begin{equation}
C=- 2 \text{Re}\left[ \frac{{C^{(1)}_g}^\ast}{|C^{(1)}_g|}\rho_{11g,01g} \right]
\end{equation}
and comes from $\ket{01}\bra{11}$ and its Hermitian conjugate (i.e., from solving the master equation with the initial condition $\rho_{01g,11g} =1$).   As in Eq.~(\ref{e12}),  the prefactor ${C^{(1)}_g}^\ast/|C^{(1)}_g|$ (with the corresponding amplitudes calculated from the non-Hermitian Hamiltonian evolution) is inserted ``by hand'' to get  $e^{-i\phi_1}$, by which the term $\rho_{11g,01g}$ needs to be multiplied when the expectation value $\bra{\Phi_\text{ideal}}\rho_f\ket{\Phi_\text{ideal}}$ is taken.  (Again, an identical contribution arising from   $\ket{10}\bra{11}$ has been included already in (\ref{a23}).) 

Finally, the terms in $D$ are
\begin{align}
D=&\rho_{11g,11g}^{(2)} + \rho_{01e_b,01e_b}^{(2)}+\rho_{01g_b,01g_b}^{(2)} \cr
&+ \frac 1 2\rho_{00g,00g}^{(2)} + \rho_{01g,01g}^{(2)} + \rho_{00e_b,00e_b}^{(2)} + \rho_{00g_b,00g_b}^{(2)} 
\end{align}
and follow from the evolution of $\ket{11}\bra{11}$, i.e., from solving the master equation with the initial condition $\rho_{11g,11g} =1$; again, in every case where there is a corresponding term with $b\to a$ its contribution has been implicitly included in (\ref{a23})).




\begin{thebibliography}{5}
\bibitem{note1} Basically, if the initial spectrum of the two photons is of the product form $f_0(\omega_a)f_0(\omega_b)$, the final state is of the form $\int_{-\infty}^\infty d\wp h(\wp,\omega_a,\omega_b) f_0(\wp)f_0(\omega_a+\omega_b-\wp)$, where, for an atom-cavity system, $h$ is a complicated function of  $\kappa$ (the cavity decay rate), $g$, the atom cavity coupling, and $\delta$, the detuning. Generally, choosing the parameters such as to minimize the entanglement (e.g, $h\to \delta(\wp-\omega_a)$) causes the nonlinear phase shift to vanish.
\bibitem{rebic2} The authors of \cite{ottaviani,rebic} did not realize this: their  model leads to $N$-atom equations that are identical to the single-atom ones except for the replacement of the single-atom coupling $g$ by $g\sqrt N$.  This is because their treatment ignores altogether the equivalent of the term $C_{ee}^{(2)}$ in Eq.~(\ref{N5}), that is, the (overwhelming, for large $N$) probability that each photon may, in fact, be absorbed by a different atom.  They appear to justify this by arguing that they work in the ``weak excitation'' limit, but this is not consistent with the fact that their results are obtained for $g\sqrt N T > 1$; as Eqs.~(\ref{M18}) show, even if the probability of each individual atom being excited is small, the probability that two atoms somewhere may be excited simultaneously becomes large as soon as $N$ is large and $g\sqrt N T \sim 1$.
\bibitem{note4} Note that some manipulation of the polarizations (to be undone afterwards) is necessary before sending the photons into the cavity, to make sure that at most only one photon of each polarization enters the cavity; this can be accomplished with polarization beam splitters and rotators.
\end{thebibliography}
 \end{document}